\RequirePackage{amsmath}    %
\documentclass[runningheads]{llncs}

\usepackage[utf8]{inputenc}
\usepackage{csquotes}
\usepackage{chngcntr}  %
\usepackage{enumitem}
\setlist{nosep,before=\vspace{.5\baselineskip},after=\vspace{.5\baselineskip}}
\usepackage{listings}  %
\usepackage{tikz}
\usetikzlibrary{patterns,arrows}
\usetikzlibrary{chains,shadows.blur}
\usepackage{pgfplots}
\pgfplotsset{compat=1.16, table/search path={include/plots}}
\usepackage{arydshln}
\usepackage{multicol}

\usepackage[prependcaption]{todonotes}

\usepackage{amsmath}

\usepackage[unicode]{hyperref}

\usepackage{cite}

\usepackage[htt]{hyphenat}

\usepackage[upgrade=true]{acro}[=v4]
\acsetup{
  use-id-as-short,
  single = true
}

\DeclareAcronym{SoK} {
    long    =   Systematization of Knowledge,
}

\DeclareAcronym{TUI} {
    long    =   Trusted User Interface,
    long-plural-form  = Trusted User Interfaces
}

\DeclareAcronym{a11y} {
    long    =   Accessibility
}

\DeclareAcronym{PoC} {
    long    =   Proof of Concept
}

\DeclareAcronym{AOSP} {
    long    =   Android Open-Source Project
}

\DeclareAcronym{DSL} {
    long    =   Domain Specific language
}

\DeclareAcronym{TCB} {
    long    =   Trusted Computing Base
}

\DeclareAcronym{RoT} {
    long    =   Root of Trust,
    long-plural-form  = Roots of Trust
}

\DeclareAcronym{TPM} {
    long    =   Trusted Platform Module
}

\DeclareAcronym{IPC} {
    long    =   Inter-process communication
}

\DeclareAcronym{TEE} {
    long    =   Trusted Execution Environment
}

\DeclareAcronym{SGX} {
    short   =   SGX,
    alt     =   Intel SGX,
    long    =   Intel Software Guard Extensions,
}

\DeclareAcronym{CSR} {
    long    =   Control and Status Registers
}

\DeclareAcronym{ROP} {
    long    =   Return-Oriented Programming
}

\DeclareAcronym{RTOS} {
    long    =   Real-Time Operating System
}

\DeclareAcronym{ISA} {
    long    =   Instruction Set Architecture
}

\DeclareAcronym{SM} {
    long    =   Secure Monitor
}

\DeclareAcronym{PMP} {
    long    =   Physical Memory Protection
}

\DeclareAcronym{IoT} {
    long    =   Internet-of-Things
}

\DeclareAcronym{CFI} {
    long    =   Control Flow Integrity
}

\DeclareAcronym{ASLR} {
    long    =   Address Space Layout Randomization
}

\DeclareAcronym{MMU} {
    long    =   Memory Management Unit
}

\DeclareAcronym{DoS} {
    long    =   Denial-of-Service
}

\DeclareAcronym{SSH} {
    long    =   Secure Shell
}

\DeclareAcronym{json} {
    long    =   JavaScript Object Notation
}

\DeclareAcronym{UART} {
    long    =   Universal Asynchronous Receiver-Transmitter
}

\DeclareAcronym{AWI} {
    long    =   Android Window Integrity
}

\DeclareAcronym{CDS} {
    long    =   Clickjacking Detection System
}

\DeclareAcronym{TZPC} {
    long    =   ARM TrustZone Protection Controller
}

\DeclareAcronym{SMC} {
    long    =   Secure Monitor Call
}

\DeclareAcronym{HPC} {
    long    =   Hardware Performance Counter
}

\DeclareAcronym{ULP} {
    long    =   ultra-low-power
}

\DeclareAcronym{RT} {
    long    =   Runtime
}

\DeclareAcronym{OS} {
    long    = Operating System
}

\DeclareAcronym{UDS} {
    long    =   Unique Device Secret
}

\DeclareAcronym{KDF} {
    long    =   Key Derivation Function
}

\DeclareAcronym{MQTT} {
    long    =   Message Queue Telemetry Transport
}

\DeclareAcronym{ECDSA} {
    long    =   Elliptic Curve Digital Signature Algorithm
}

\DeclareAcronym{LoC} {
    long    =   Lines of Code
}

\DeclareAcronym{DICE} {
    long    =   Device Identifier Composition Engine
}

\DeclareAcronym{ecall} {
    long    =   environment call
}

\DeclareAcronym{ePMP} {
    long    =   Enhanced Physical Memory Protection
}

\DeclareAcronym{API} {
    long    =   Application Programming Interface,
}

\DeclareAcronym{M-Mode} {
    long    =   Machine Mode
}

\DeclareAcronym{S-Mode} {
    long    =   Supervisor Mode
}

\DeclareAcronym{U-Mode} {
    long    =   User Mode
}

\DeclareAcronym{H-Mode} {
    long    =   Hypervisor Mode
}

\DeclareAcronym{eapp} {
    long    =   enclave application,
    long-plural-form    =   enclave applications
}

\DeclareAcronym{AES} {
    long    =   Advanced Encryption Standard
}

\newcommand{\PROJNAMETITLE}{R5Detect}
\newcommand{\PROJNAME}{\texttt{R5Detect}}

\hypersetup{hidelinks}

\begin{document}

\title{\PROJNAMETITLE{}: Detecting Control-Flow Attacks from Standard RISC-V Enclaves}

\author{Davide Bove
\and
Lukas Panzer}
\authorrunning{D. Bove \& L. Panzer}

\institute{Friedrich-Alexander-Universität Erlangen-Nürnberg, Germany
\email{\{davide.bove,lukas.panzer\}@fau.de}
}

\maketitle              %

\vspace{-\baselineskip}

\begin{abstract}
Embedded and Internet-of-Things (IoT) devices are ubiquitous today, and the uprising of several botnets based on them (e.g., Mirai, Ripple20) raises issues about the security of such devices.
Especially low-power devices often lack support for modern system security measures, such as stack integrity, Non-eXecutable bits or strong cryptography.

In this work, we present \PROJNAME{}, a security monitoring software that detects and prevents control-flow attacks on unmodified RISC-V standard architectures.
With a novel combination of different protection techniques, it can run on embedded and low-power IoT devices, which may lack proper security features.
\PROJNAME{} implements a memory-pro\-tec\-ted shadow stack to prevent runtime modifications, as well as a heuristics detection based on Hardware Performance Counters to detect control-flow integrity violations.
Our results prove that regular software can be protected against different degrees of control-flow manipulations with an average performance overhead of below 5~\%.
We implement and evaluate \PROJNAME{} on standard low-power RISC-V devices and show that such security features can be effectively used with minimal hardware support.

\end{abstract}

\section{Introduction}

Today there are many internet-connected devices on the market that, once deployed, are difficult to manage remotely by a vendor.
Some of these are so-called \acf{IoT} consumer devices, such as smart fridges, but there are also embedded systems that are used in critical infrastructure deployments.
The more devices there are, the more interesting they become for potential attackers, which may discover a vulnerability in one device and compromise millions of targets, as seen in \ac{IoT}-based botnets like Mirai~\cite{DBLP:conf/uss/AntonakakisABBB17} or Mozi~\cite{DBLP:journals/ijis/TuQZCXH22}.

One problem of \ac{IoT} and embedded devices is that the majority of devices on the market are mostly designed to be low-power and cheap customer devices.
Therefore, low-end hardware is often used or critical security features that may require more processing power, such as strong encryption, are not available~\cite{DBLP:journals/micro/YangBS17}.
An increasing number of \ac{IoT} devices is being deployed in consumer homes, and a vulnerable device may compromise the security of several households~\cite{DBLP:conf/sp/AlrawiLAM19}.
It is therefore desirable to have both affordable devices and strong security measures against attacks.
Both complex software like a Linux kernel and more lightweight \ac{OS} kernels like FreeRTOS are prone to vulnerabilities and attacks~\cite{orikarliner2018}.
In this work, we are looking at detecting firmware and software manipulation attacks at runtime to prevent devices from going rogue.

There are several ways to detect runtime manipulations of the firmware or software on a device.
Mostly we look at control flow manipulations, where an attacker may change the order of existing instructions or introduce new ones using vulnerabilities in the deployed software.
On most architectures, there are so-called \acfp{HPC} that count specific events occurring at runtime.
This information can be used to detect potential modifications~\cite{DBLP:conf/iccad/WangKMK15}.
During execution, the performance of a software is measured and continuously compared to the expected output.
Another method is to observe the execution of the software and check for integrity violations of the control flow.
We are the first to explore both methods and to combine them in a single monitoring system that will detect \acf{CFI} violations and unexpected software behavior on low-power \ac{IoT} and embedded devices based on the RISC-V architecture.

Therefore, we created \PROJNAME{}, a security monitoring software for low-power devices that combines the following \textbf{contributions}:

\begin{itemize}
    \item We implement \ac{CFI} security checks with a Shadow Stack implementation using standard RISC-V features.
    \item We implement a monitoring heuristics based on \aclp{HPC}.
    \item We improve the overall security using built-in defenses to protect \PROJNAME{} against strong attackers.
    \item We test and successfully deploy \PROJNAME{} on an FPGA and a low-power RISC-V board.
    We also quantify both the effectiveness and the performance of the different detection methods.
\end{itemize}

We provide some additional background information for the reader in \autoref{r5detect:background}.
In \autoref{r5detect:threat-model} we describe the security constraints of our approaches.
We present the two separate security monitoring approaches used in \PROJNAME{}, namely our implementation of a CFI monitor in \autoref{r5detect:cfi} and the HPC monitoring in \autoref{r5detect:hpc}.
Regarding existing related work (see \autoref{r5detect:related-work}), we discuss our results in \autoref{r5detect:discussion} and summarize our results in \autoref{r5detect:conclusion}.

\section{Background} \label{r5detect:background}

In order to achieve a robust monitoring on a mostly untrusted device, we require hardware support for specific security features.
For our implementation of \PROJNAME{}, memory protection features and a separation of hardware resources in different privilege levels are required.
This is found on several platforms and processors, but both features are also natively covered by the RISC-V standard architecture.
RISC-V is a license-free and open-standard instruction set architecture.
Therefore, we can easily implement our monitoring setup with unmodified commodity hardware boards.

We implement \PROJNAME{} on top of MultiZone, a \ac{TEE} framework for RISC-V, to guarantee that the monitoring is safe from attackers.
MultiZone uses standard features available on most RISC-V devices to protect software from manipulations even in the presence of a compromised \ac{OS} kernel.

\subsection{RISC-V Security}
In this work, we use several standardized features of the RISC-V architecture.
Since most low-power devices on the market only have limited CPU features, most devices cannot run a full Linux kernel and often rely on more adapted \ac{RTOS}.
\PROJNAME{} targets such devices and only requires two different privilege levels, which is recommended in the RISC-V Instruction Set Architecture (ISA) for \enquote{Secure embedded systems}~\cite{Waterman:riscv-priv-20211203}.
Our solution also relies on memory protection, as described below.

\subsubsection{Privilege Levels}
Comparable to \enquote{Rings} in x86 and \enquote{Exception Levels} in ARM, the RISC-V Privileged Architecture defines different privilege levels in which software can run.
On a platform level, we require at least two levels to achieve some secure separation.
The mandatory Machine Mode (M-Mode) has the highest privileges and is always present on compliant RISC-V devices.
In our solution, the most important software is called \enquote{Secure Monitor} (SM) and runs in M-Mode, such that any compromise of the \ac{OS} might not directly affect the security of the whole system. 
The second mode is the User Mode (U-Mode), which is intended for running the operating system with its user apps.
In the specification there are even more modes, such as the Hypervisor Mode (H-Mode) or the Supervisor Mode (S-Mode) which are intended for running more complex systems using virtualized environments or a Linux kernel.
Since our focus is on low-power and embedded devices, we do not require these modes to be implemented.

\subsubsection{Physical Memory Protection}
In order to manage memory access from different privilege modes, RISC-V includes \acf{PMP}, a hardware-based security feature.
PMP regulates the access to specific memory regions from lower-privileged S-Mode and U-Mode.
With special registers, M-Mode software can set rules for memory that can be any combination of read, write and execute permissions.
There is also the \textit{Lock Bit}, which allows locking such a rule until the next device reboot.
The Lock Bit enforces rules for all modes, and locked PMP entries cannot be \enquote{unlocked} during runtime (not even in M-Mode) until the device is restarted.

\subsection{Trusted Execution Environment}
A \acl{TEE} is a processor feature that offers some kind of isolation inside the hardware for processing data.
The goal is to have a secure execution of specific applications even in the presence of a compromised \ac{OS}.
Such applications, called trusted apps or trustlets, are responsible for processing security-critical operations, such as encryption or authentication.
Often they also provide services that handle sensitive user data, which may need extra protection from strong attackers.
These isolated environments are often called \textit{enclaves}, while this work will reference them as \textbf{zones}.
Zones run in U-Mode and have only access to other zones by shared memory spaces, which are defined and assigned by the Secure Monitor.

\subsection{OpenMZ}
\PROJNAME{} is built on top of OpenMZ, an open-source implementation of MultiZone Security~\cite{openmz20}.
MultiZone is a \ac{TEE} framework for RISC-V devices, which allows different applications to run in isolated environments.
Our solution can monitor such an environment as a privileged entity, while still being protected from attacks itself.

OpenMZ splits the system into different \textit{zones}, which can contain up to 8 memory regions.
Memory isolation between zones is implemented using \ac{PMP}.
OpenMZ is used as a Secure Monitor, which is the software running in M-Mode and which manages the context switches between zones.
The framework implements the MultiZone \ac{API} and has a code base of 758 \ac{LoC}.
The compiled binary size is approximately 4 KiB, which makes it a suitable \ac{TEE} for embedded and resource-constrained systems.

\section{Threat Model} \label{r5detect:threat-model}
\enquote{\ac{IoT}} describes a vast amount of different devices, so without a precise definition, it might be unclear why some decisions were made.
There is no consensus how to define low-power or ultra-low-power devices, but when talking about \ac{IoT} devices we mean systems that are \enquote{equipped with sensors, actuators, computers, and network connectivity}~\cite{DBLP:journals/micro/YangBS17}.
Most devices we target have a flash memory chip and limited system resources, such that they cannot run fully-featured operating systems.
Our two test boards, for example, are not suited to run a Linux Kernel, which is why our test applications are bare-metal apps or rely on a small-scale \ac{OS} like FreeRTOS, a \acl{RTOS}.
In our experience, a lot of these devices do also not support many security features, such as Secure Boot or \ac{ASLR}.
Moreover, we do not require a \acf{MMU}, which we found is often not available.
Most \ac{RTOS} for these devices operate on the memory model of \enquote{logical address space equals physical address space} and do not support address space translations.
Especially with RISC-V, most low-power devices support only one and a maximum of two CPU privilege levels (see more in \autoref{r5detect:background}).
Embedded devices share similar limitations and are often a central part of \ac{IoT} systems.

Before describing the technical details of our defenses, we define our general threat model.
On our target devices, we run \PROJNAME{} in the privileged \ac{M-Mode}, while the monitoring code and other applications (bare-metal, FreeRTOS app) run in a zone in \ac{U-Mode}.
Since the monitoring is running in the unprivileged \ac{U-Mode}, even a compromise of it would prevent further escalation to the \acl{SM}.

We assume an unprivileged \ac{U-Mode} attacker, which means that the \acl{SM} is uncompromised and the attacker may exploit one or all of the user zones, except the monitoring zone.
The attacker may modify stack contents of each individual zone, use memory-based vulnerabilities to manipulate the control flow of apps, and ultimately take over a zone.
The attacker may also introduce own code, eventually circumventing the detection methods of \PROJNAME{}.

Our work assumes that the integrity of the \acl{SM} is safe and there are Secure Boot mechanisms, such that manipulations of the \acl{SM} or the firmware are detected and prevented. 
The monitoring zone does not leak sensitive information, such as stack contents or register values. 
If the monitoring zone is compromised, all user-mode apps can be considered compromised, but not the \acl{SM}.

\ac{DoS} attacks are considered in the design.
Attackers may try to stop or tamper with the monitoring, or completely shut down the device.
While unprivileged apps may perform actions that damage the board or manipulate computing states, attackers cannot block or starve out the monitoring, as the \acl{SM} includes an own scheduler, which prevents zones from blocking the execution of other zones.
A \ac{DoS} attack may only be possible if the trusted monitoring zone is directly compromised (out of scope for this work), in which case the detection results become unreliable.

Furthermore, we focus on software-based attacks, meaning that hardware attacks are out of scope.
This includes fault-injection~\cite{DBLP:conf/date/WernerSUM19, DBLP:journals/tches/NashimotoSUH22} and side-channel attacks~\cite{DBLP:conf/dac/MulderGH19, DBLP:journals/corr/abs-2106-08877}, which in general are too powerful to protect against, especially on low-power devices as described in this chapter.
Side-channels may reveal the presence of the monitoring, but not manipulate its functioning.
While \textit{all} attacks need to be considered for a secure platform, we consider the overall impact of a physical attacker on \ac{IoT} devices to be quite low compared to large-scale software-based vulnerabilities and attacks.

\section{Control-Flow-Integrity Using a Shadow Stack} \label{r5detect:cfi}
Many software-based attacks exploit the control flow of vulnerable apps to change or delay the execution of specific functions.
Therefore, these attacks are called control-flow integrity attacks.
Most modern systems already implement protections against \ac{CFI} attacks, such as non-executable data segments, stack canaries \cite{DBLP:conf/uss/Cowan98} or runtime randomization \cite{DBLP:conf/srds/XuKI03}.

Unfortunately, some of these protections can be circumvented, for example by reusing code segments in \ac{ROP} attacks \cite{DBLP:journals/ieeesp/PrandiniR12,DBLP:conf/raid/CloostersPWDJSD22}.
Randomized code layouts and stack canaries can also be predicted and circumvented using side-channels~\cite{DBLP:journals/ieeesp/PincusB04,DBLP:conf/ccs/ShachamPPGMB04,DBLP:journals/tissec/AbadiBEL09}.
\ac{CFI} defenses may offer a robust protection specifically against \ac{ROP} attacks and buffer overflow attacks.
Furthermore, \ac{CFI} protections can be implemented in software, or one can extend and customize the hardware to support them \cite{DBLP:conf/date/DeBGJ19}.
Since we try to make such protections available for many devices, we implemented a software-based solution which targets the standard RISC-V \ac{ISA}.

\subsection{Setup} \label{r5detect:cfi-setup}
Before designing our protection, we consider the threat model that we want to protect against (see \autoref{r5detect:threat-model}).
We assume an attacker that may control main memory data contents of the executing user application.
This may be done using vulnerabilities in the program, stack manipulation techniques etc.
The goal of the attacker is to redirect the execution, such that instructions controlled by the attacker may be executed instead of the original instructions.
This threat model has only three limitations that need to be guaranteed for our solution:

\begin{enumerate}
    \item The \texttt{.text} segment of the program, the actual code executed on the system, is not writable.
    \item Data segments of the program are marked as non-executable.
    \item The attacker may not directly manipulate registers.
\end{enumerate}

From experience with actual devices, most of the above points are already covered by the processor.
Therefore, we do not consider these to be too limiting for real-world use cases.
(1) prevents an attacker from manipulating or circumventing our security checks.
As we rely on the integrity of our binary instrumentation, this is a crucial requirement.
Since shared libraries on embedded devices are often statically linked into a binary, the code segment does not need to be writable.
(2) prevents an attacker from transforming data into instructions such that they can be executed.
An attacker may manipulate any data on the stack, but not change the execution flags.
Since our protection uses some registers to do the security checks, (3) must be guaranteed to some extent for our solution to work.
This limitation may also be restricted to certain registers, depending on the number of available registers on the system.

\subsection{Implementation}
In general, software running on a machine uses branching and jumps to redirect the execution of the code to different functions.
There are direct jumps, that specify the next address to execute, and there are \textit{indirect} jumps, that specify a location where the target address is stored (e.g., a register).
A problem with indirect jumps is that they can target any executable address, as register contents may be changed during execution.

A similar challenge exists with return instructions.
RISC-V encodes a \texttt{ret} instruction as an indirect jump \texttt{jalr x0, ra, 0}.
In this instruction, the return address (e.g., the caller function) saved in register \texttt{ra} may be saved on the stack, where an attack might manipulate it.
This manipulation of return addresses is the basis for \ac{ROP} attacks.
Furthermore, while return instructions on other architectures target one fixed address (the next instruction of the caller function), the indirect jump on RISC-V may lead to multiple targets, as register contents may be changed or manipulated.
Therefore, we cannot check the validity of a return \enquote{jump} before it is executed.

So, our goal here is to prevent such arbitrary jumps to counter \ac{CFI} attacks.
The implementation of the \ac{CFI} monitoring requires instrumentation of the program's binary.
We present two approaches to do security checks for the different cases described above.

\subsubsection{Indirect jumps}
RISC-V has two instructions for indirect jumps: \texttt{jalr} (jump and link register) and \texttt{jr} (jump register).
In order to provide a protection, our monitor needs to check if the target addresses are in an allowed set of addresses.
For this, we use assembly \textbf{labels} that are injected into a binary after compilation (binary rewriting) and define which locations an indirect jump may target inside a program, derived from the call graph of the application.
At runtime, we check if the target has the correct label (\textit{forward check}) before executing a jump.
If the label does not match, the jump is not executed.

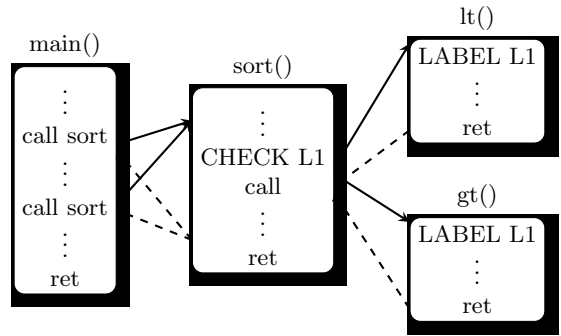
\begin{figure}[htbp]
		\begin{center}
			\begin{tikzpicture}[transform shape, scale=1.0,box/.style = {draw,rounded corners,blur shadow,fill=white, align=center}]
				
				\node[box,label={main()}] (main) at (0, 0)   {$\vdots$\\ call sort\\$\vdots$\\ call sort\\$\vdots$\\ret};    
				
				\node[box,label={sort()},right=of main] (sort)   {$\vdots$\\ CHECK L1 \\ call \\$\vdots$\\ret};    
				
				\node[box,label={lt()}, above right=-0.75cm and 1cm of sort] (lt)   {LABEL L1\\$\vdots$\\ ret}; 
				
				\node[box,label={gt()}, below right=-0.75cm and 1cm of sort] (gt)   {LABEL L1\\$\vdots$\\ ret};

				\draw[->,thick,>=stealth] (main.35) --  (sort.140);
				\draw[->,thick,>=stealth] (main.335) --  (sort.140);	
				\draw[->,thick,>=stealth] (sort.10) --  (lt.145);	
				\draw[->,thick,>=stealth] (sort.5) --  (gt.145);
				
				\draw[->, dashed,thick] (lt.210) --  (sort.355);
				\draw[->, dashed,thick] (gt.210) --  (sort.350);
				\draw[->, dashed,thick] (sort.220) --  (main.30);
				\draw[->, dashed,thick] (sort.220) --  (main.330);
			\end{tikzpicture}
		\end{center}
	\caption{Example program with indirect jumps and forward checks using labels. The \texttt{sort} function has two valid sub-functions (\texttt{lt} and \texttt{gt}), which are determined and annotated at compile time. }
    \label{fig:r5detect-cfi-example1}
\end{figure}

An example of this is found in \autoref{fig:r5detect-cfi-example1}. The target address of the \texttt{sort} function is passed by parameter, but the only valid targets are the functions \texttt{lt()} and \texttt{gt()}.
The target addresses are labeled, and these labels are checked before the jump in the sort function.

\subsubsection{Return instructions}
As return addresses are saved on the stack, which may be manipulated by overflow and stack corruption attacks, we need to move them out of reach of a program's execution.
Our solution implements a \textbf{Shadow Stack}, a data structure used to store all the return addresses.
For this, every instruction that saves a return address on the stack is replaced by a call to the Shadow Stack.
The Shadow Stack is managed by the \acl{SM} and located in the isolated monitoring \textit{zone}. %
This zone may contain more than one Shadow Stack: one for every zone or multiple ones for a zone.
Whenever a program pushes or pulls from the stack, an \acf{ecall} instruction is executed that moves the execution to the \acl{SM}.
The \acl{SM} handles the request and loads or saves the previously saved stack address into the \texttt{ra} register.
Moreover, checks are included to prevent stack overflows and underflows.
This does introduce a minor performance overhead (see next section), but does not interfere with the correct execution of applications.

\subsection{Evaluation}
In order to test the effectiveness of our approach, we tested \PROJNAME{} on two hardware devices.
We used the HiFive1 Rev B with the native RISC-V chip \textit{FE310} with a clock speed of 320 MHz.
As an alternative with more RAM, we used the Artix-7 35T FPGA with the Hex Five \textit{X300 RV32ACIMU} bit stream and a clock speed of 65 MHz.

We evaluate the security of our approach by checking our implementation against the security guarantees and assumptions of our threat model (see \autoref{r5detect:cfi-setup}).
In addition, we used generic buffer overflow attacks to simulate an attack and evaluate how \PROJNAME{} may detect such attacks.
Furthermore, we run benchmarks to evaluate the impact on the system as well as the general performance of our approach.

\subsubsection{Security}
As shown in related work, correctly implementing the Shadow Stack should yield no potential vulnerabilities for the chosen threat model~\cite{DBLP:journals/tissec/AbadiBEL09}. %
The implementation checks all indirect jumps, and these checks can not be circumvented as long as the \acl{SM} and the Monitoring Zone are safe.
Attacks that follow a regular \enquote{legal} control flow may still go undetected, so our approach does not necessarily detect other types of vulnerabilities.
The security of our \ac{CFI} monitoring is based on three pillars:

\begin{enumerate}
    \item \textbf{Memory protection}: This is achieved using \ac{PMP}. Executable code is made read-only, and the data segments are non-executable, so there is no possibility for an attacker to inject new code.
    
    \item \textbf{Labels}: Since our approach annotates compiled software binaries, the label IDs in the program must be unique.
    This property is statically checked by the \acl{SM} before code execution, such that an attacker might not introduce own labels during runtime.
    
    \item \textbf{Jump checking}: With our forward and backward jump checks, we make sure that jump instructions may only occur for the defined \enquote{legal} paths. If the attacker succeeds to inject a label into the writable memory space, the jump will succeed as it is considered legal to the \acl{SM}.
    In this case, (1) will still ensure that code in writable segments may not be executed.
    For the Shadow Stack, an application never has direct access to it, as only the \acl{SM} may read or write to it. 
    Therefore, stack vulnerabilities do not affect the security of the system. 
\end{enumerate}

For our security testing, we used an exemplary vulnerable binary that, using \texttt{memcpy}, allows stack buffer overflows.
We implemented a total of three proof-of-concept attacks that target this vulnerable binary.

Our first attack tries to redirect the execution of the binary by overwriting the return address on the stack.
Such an attack is easily detected by \PROJNAME{}, but it may also be prevented using a shadow stack implementation managed by the application itself, as offered by some compilers (e.g., Clang's ShadowCallStack\footnote{\url{https://clang.llvm.org/docs/ShadowCallStack.html}}).

The second attack assumes an attacker that knows about the shadow stack, its position in memory and the stack pointer.
In a real-world scenario, the presence of our solution may be discovered, either by trial-and-error, luck or through side-channel information leaks.
In this scenario, the attacker tries to overwrite a pointer address and a local variable to manipulate a value on the shadow stack.
This attack would be successful if the shadow stack would be managed by the application (e.g., using the above \textit{ShadowCallStack}).
Since \ac{PMP} protects our shadow stack implementation from any access outside the app's zone, the system throws a \enquote{Load access fault} exception and prevents the manipulation.

The third attack overwrites the address of an indirect jump.
We assume the vulnerable binary has the following call graph, with \texttt{func1} and \texttt{func2} as functions that call other functions (labeled \texttt{TARGET*}) based on a conditional parameter:
\begin{align*}
\texttt{func1 $\rightarrow$ [TARGET1 | TARGET2]} \\
\texttt{func2 $\rightarrow$ [TARGET2 | TARGET3]}
\end{align*}

The attack replaces the jump target on the stack of \texttt{func1} with the address of \texttt{TARGET3} of the program.
\begin{align*}
\texttt{func1 $\rightarrow$ TARGET3}
\end{align*}

While \texttt{TARGET3} is a \enquote{legal} label, the control flow above does not exist in the call graph.
Therefore, it is an illegal control flow and \PROJNAME{} prevents the jump.
On the other hand, replacing \texttt{TARGET1} with \texttt{TARGET2} on the stack for \texttt{func1} may not be detected, as it is a legal path.
If an attacker injects a legal (e.g., a duplicate) label in the data segment of the memory and manipulates the stack accordingly, \ac{PMP} still prevents the execution, as all data segments are marked as non-executable.

\subsubsection{Performance}
In order to measure the performance of our approach and the overhead caused, we run several benchmarks from the BEEBS suite~\cite{DBLP:journals/corr/PallisterHB13}.
The suite is designed to measure the energy consumption of embedded devices, which is not relevant for our target devices.
But it also combines a decent range of different benchmarks focused on specific processor features, such as branching, memory access and more, which makes it suitable for our purposes.
It is one of very few frameworks that can be easily ported and compiled to the RISC-V architecture, which was a major factor for choosing it.
In addition, there is related work using BEEBS, so we can compare our results more easily.

We evaluate the runtime overhead by running the benchmarks with and without the \ac{CFI} instrumentation.
The benchmarks run in a single zone with default settings and 128 iterations.
We compiled the benchmarks with \texttt{gcc} and optimization level \texttt{O3}.
For timing the benchmarks, we use the real-time clock \texttt{mtime} and the internal BEEBS functions \texttt{start\_trigger} and \texttt{stop\_trigger}.

\begin{figure}[htbp]
	\begin{center}
		\hspace*{-15pt}
		\begin{tikzpicture}
			\begin{axis}[
				scale only axis=true,
  				height=4.5cm,
				enlarge x limits=0.03,
				every axis plot post/.style={/pgf/number format/fixed},
				ybar=1pt,
				ylabel=runtime overhead in percent, 
				bar width=5pt,
				x=0.24cm,
                ytick={50, 100, 200, 300, 400},
				ymin=0,
                ymajorgrids,
                grid style={dashed,gray!25},
				axis on top,
                scale=0.7,
				xtick=data,
				xticklabel style={rotate=90, xshift=0pt, font=\tiny},
				xticklabels from table={cfi-runtime-overhead.txt}{Benchmark},
				visualization depends on=rawy\as\rawy, %
				after end axis/.code={ %
					\draw [ultra thick, white, decoration={snake, amplitude=2pt}, decorate] (rel axis cs:0,1.05) -- (rel axis cs:1,1.05);
				},
				nodes near coords, every node near coord/.append style={
					anchor=west,
					rotate=90,
					font=\tiny},
				axis lines*=left,
				clip=false
				]
				\addplot[draw = blue, 
				fill = blue!50!white,
				/pgf/number format/read comma as period] table [x=x, y=y] {cfi-runtime-overhead.txt};
			\end{axis}
		\end{tikzpicture}
		\caption{Run-time overhead of different BEEBS benchmarks for the Shadow Stack implementation compared to unmodified benchmarks}
		\label{fig:r5detect-cfi-runtime-overhead}
	\end{center}
\end{figure}
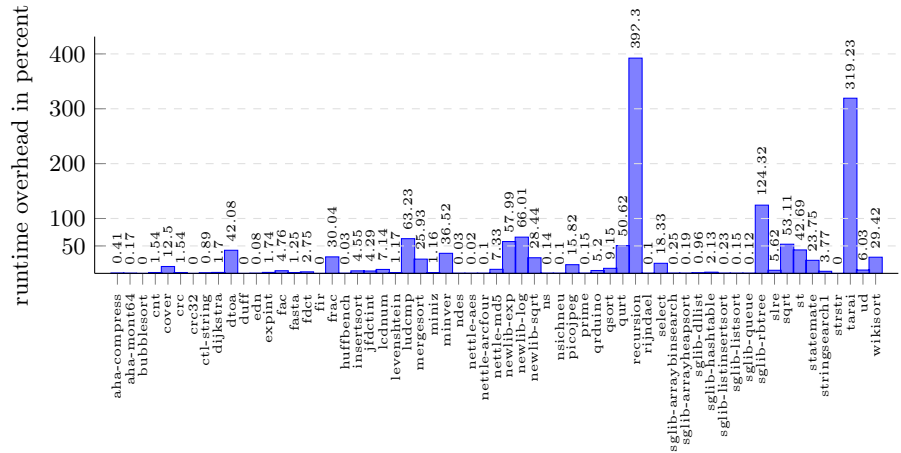

The results of our measurements can be seen in \autoref{fig:r5detect-cfi-runtime-overhead}.
Our binary label checking approach combined with the Shadow Stack causes a runtime overhead of $<$ 5~\% for over 50~\% of the benchmarks.
Individual benchmarks with many indirect jumps, such as \texttt{mergesort}, as well as recursive functions with many calls to small methods, such as \texttt{recursion} and \texttt{tarai}, have a significant overhead to their runtime.
These results \textbf{are expected}, as software with many function calls causes more jump checks.
On the other hand, recursion and excessive function calls are mostly avoided in embedded systems, so not many real-world applications achieve such high runtime overhead~\cite{DBLP:conf/ecrts/WallsBBSOW19}.
As a solution, the label checking can be disabled for single functions or after a certain depth of sub-functions is reached. 

Since labels are injected into the binary files, the code segments increased with the number of functions. %
Overall, the binary size increases by 7.21~\% on average with a median of 4.32~\%.
Again, the benchmarks with more functions tend to get bigger because of the instrumentation method.
Moreover, being optimized for embedded systems and with compiler optimizations enabled, most benchmarks are minimal binaries (5-50 KB depending on benchmark and flags), such that even few bytes lead to a high percentage increase.

The results indicate that the security goals can be achieved with our \ac{CFI} method with relatively low performance overhead.
We discuss the limitations of this approach in \autoref{r5detect:discussion}, while the next section will present our method for anomaly detection.

\section{Hardware Performance Counters to Monitor Suspicious System Activity} \label{r5detect:hpc}
Since the above \ac{CFI} defense has some drawbacks (see \autoref{r5detect:discussion}), we implemented a separate detection method based on \aclp{HPC}.
\acp{HPC} are special registers that allow to count specific internal processor events.
There are plenty of events that can be tracked, but the counters are limited, such that not all events can be monitored simultaneously.
These counters may also be used for integrity checks of programs \cite{DBLP:conf/ccs/MaloneZK11} or for anomaly detection \cite{garciaserrano15}.

Observable events depend on the actual hardware implementation and may include:

\begin{itemize}
    \item cache misses,
    \item branches and jumps,
    \item arithmetic operations
\end{itemize}

A monitoring software may use these and other events to detect unexpected system behavior \cite{DBLP:conf/ccs/MaloneZK11}.
The RISC-V \ac{ISA} defines its own set of machine-level performance counters, implemented as so-called \acfp{CSR}~\cite{Waterman:riscv-priv-20211203}.
Every higher-privileged mode (e.g., M-, H-, and S-Mode) has their own set of \acp{CSR}, which include:

\begin{itemize}
    \item the number of processor cycles: \texttt{mcycle}
    \item the number of \textit{retired} (executed) instructions: \texttt{minstret}
    \item a high-resolution real-time counter: \texttt{mtime}
\end{itemize}

Processors may use more unprivileged registers for event counting, so the actual number of usable \acp{HPC} depends on the implementation of the hardware.

\subsection{Setup}
In order to profile and monitor different zones, we adapted our \acl{SM} to save and manage \acp{HPC} for every zone individually.
We assume the same threat model as before (see \autoref{r5detect:cfi-setup}).
Therefore, every zone has its own active counters, which are saved when a context switch happens (to another zone or the \acl{SM}).
The monitoring runs on the same privilege level than other zones (U-Mode), but has added privileges, which means it can access the counters of all the other zones.
This is achieved by adapting the \ac{PMP} settings for the privileged zone during execution.
We also adapted the \acl{SM} scheduler such that the monitoring zone is executed regularly and may not starve out.
Since U-Mode software has no access to privileged \acp{CSR}, only the \acl{SM} can access the performance counters.

For the detection, we use a signature-based approach where we profile the performance of counters in an \enquote{offline} learning phase with several executions.
During the actual execution, the \acl{SM} is notified when a violation is detected.
The actual reaction to the violation can differ depending on the use-case.

\subsection{Implementation}
Monitoring with \acp{HPC} requires hardware registers to count the events.
RISC-V defines up to 29 different general-purpose registers (31 registers excluding \texttt{x1/ra} for the return address and \texttt{x2/sp} for the stack pointer)~\cite{Waterman:riscv-2019121},
but most devices only implement a fraction of these to be used with \ac{HPC}.
An overview of available \ac{HPC} can be usually found in the data sheet of the devices.
For our test boards, 23 different events can be counted~\cite{fe310manual2022}. %

Since \acp{HPC} are a hardware feature, they have no awareness of the system or the OS running on the device.
We need to add the context of current zones and privilege levels in the implementation of the \acl{SM}.
The data structure for \textit{zones} was extended such that it manages the current value of a specific \ac{HPC}.
Whenever a zone is entered or exited, we respectively load or save the current value of the counters.
Therefore, every zone is assigned its own \ac{HPC} counters, and we can differentiate and compare them easily.

In the offline phase, we collect and create a unique signature with expected \acp{HPC} for a specific application, which is known to the \acl{SM} when the device is running.
This also means that an attacker has no access to the signature (stored in the monitor zone) and therefore cannot anticipate if and which manipulations might be detected.
After the monitor has collected enough \ac{HPC} values for every zone, it matches the values with the static signature for every application and may raise an alarm if there is a mismatch.
The reaction to this event is left to the vendor and developer of the system and is out of scope.

\subsection{Evaluation}
For our experiment, we selected six \ac{HPC} values and two test binaries from the BEEBS benchmark suite.

\subsubsection{Effectiveness}
The binary \textit{picojpeg} is a JPEG decoder for low-power devices and microcontrollers\footnote{\url{https://github.com/richgel999/picojpeg}}.
The binary \textit{nettle-aes} is an \ac{AES} implementation and part of the cryptographic library Nettle\footnote{\url{https://www.lysator.liu.se/~nisse/nettle/nettle.html}}.
Because of their implementations, the two programs perform very differently, which is why we selected them for our evaluation.
In order to test the security, we simulate code manipulations by modifying the benchmark binaries.
In addition to their normal execution, we inject code to achieve a control-flow manipulation and to simulate an attack.
The modifications are as follows:

\begin{itemize}
    \item \texttt{picojpeg-mod1} changes an AND to an OR, modifying the output while still retaining a valid call graph.
    \item \texttt{picojpeg-mod2} introduces a new function that calls a number of different sub-functions.
        This simulates a buffer overflow and a \ac{ROP} attack, invalidating the original call graph.
    
    \item \texttt{nettle-aes-mod1} changes the key size for the \ac{AES} encryption from 256 to 192, which retains a valid call graph.
    \item \texttt{nettle-aes-mod2} is modified in the same way as above and introduces (and calls) a new function.
        Again, this simulates a buffer overflow and a \ac{ROP} attack.
\end{itemize}

In addition to that, we experimented with different inputs to show how much they would influence the \ac{HPC} values for the unmodified binary.
For \textit{picojpeg}, we selected two JPEG files with the same resolution, but different quality (and therefore different sizes).
For the \textit{nettle-aes} binary we changed the input file for encryption (\texttt{inp1}) as well as the encryption key (\texttt{inp2}).
While we experimented with all available \acp{HPC}, only six of them were suited to be used for our test binaries.
The resulting \ac{HPC} value differences (deviations from the unmodified binary averaged over 1000 executions) are shown in \autoref{tbl:r5detect-hpc-benchmarks}.

\begin{table}[tb]
    \scriptsize
	\centering
	\begin{tabular}{|c|cccccc|} \hline
    	\multicolumn{1}{|l|}{} & \multicolumn{6}{c|}{Deviation in \%}        \\
        Binary                & INT  & JAL   & CB   & MIO  & PFE    & BDM   \\ \hline
        picojpeg-inp2         & 42.5 & 24.4  & 31.0 & 45.6 & 111.8  & 31.0  \\
        picojpeg-inp3         & 32.2 & 35.3  & 18.0 & 33.7 & 120.2  & 12.4  \\ \hdashline
        picojpeg-mod1         & 12.4 & 33.0  & 15.6 & 3.3  & 18.9   & 4.0   \\
        picojpeg-mod2         & 14.9 & 148.3 & 2.2  & 24.6 & 2040.5 & 5.9   \\ \hline
        nettle-aes-inp2       & 0.1  & 0.0   & 0.2  & 0.1  & 0.2    & 0.4   \\
        nettle-aes-inp3       & 0.2  & 0.2   & 0.0  & 0.1  & 0.1    & 0.3   \\ \hdashline
        nettle-aes-mod1       & 1.8  & 21.4  & 10.8 & 0.4  & 0.1    & 4.3   \\
        nettle-aes-mod2       & 0.4  & 95.2  & 3.0  & 0.1  & 0.7    & 4.5   \\ \hline
    \end{tabular}
	
	\caption{Deviation of HPC values from default run for different modifications and events: 
    	integer arith inst retired~(INT),
    	JAL instruction retired~(JAL),
    	cond branch retired~(CB),
    	memory-mapped I/O access~(MIO),
    	pipeline flush other events~(PFE),
    	branch direction misprediction~(BDM).
	}
	\label{tbl:r5detect-hpc-benchmarks}
\end{table}

The results indicate that some events perform better at detecting control flow modifications than others, e.g., \texttt{PFE} for \texttt{picojpeg-mod2} and \texttt{JAL} for \texttt{nettle-aes-mod2}.
We also see why we chose these binaries for the experiment.
The control flow of the JPEG decoder greatly depends on the input, which is why we have such variations for different inputs.
This is not true for \ac{AES}, which always executes the same number of encryption \enquote{rounds} for different input files and keys.

The next insight is that for every modification, not every \ac{HPC} performs well.
For \ac{AES}, as in its implementation most instructions happen in a loop, the number of executed instructions is more significant than other factors.
Therefore, minor modifications which do not breach the allowed call graph (\texttt{*-mod1}) are, in general, more difficult to detect.
Furthermore, selection of appropriate \acp{HPC} greatly depends on the executable, which is why you need to use more different inputs to \enquote{train} the system better.
We can see with the example of \textit{picojpeg} why that is not feasible.
As an infinite number of files can be legal JPEG files, you cannot cover all inputs.
Therefore, training data will always be missing some valid execution paths for the program.

A solution to this problem might be to build dependencies between \acp{HPC}~\cite{DBLP:conf/ccs/MaloneZK11}.
In our case, the HiFive1 board only supports two active \acp{HPC}, which does not allow building robust correlations between events.
In general, this is a severe limitation of our experiment, which we discuss in \autoref{r5detect:conclusion}.

\subsubsection{Performance}
The \ac{HPC} detection approach requires a separate \textit{zone} in the MultiZone system.
In our experiment, the monitoring zone takes 32 KiB of ROM.
During execution, the zone occupies 1 KiB of main memory for stack and heap, which is approximately 8.33~\% for the HiFive1 Rev B.
On the Arty A7 board, the total memory overhead is 1.67~\%, due to its bigger available memory.

The runtime performance depends on the chosen scheduler strategy.
We selected an interval of 100 ms for our round-robin scheduler, which leads to a performance overhead of 0.1~\% compared to non-monitored applications.
Since our monitoring zone finishes its execution before this interval, the overhead is realistically below the above number.
We found that the memory bus of the HiFive1 Rev B is significantly slower than comparable boards (e.g., the Arty A7), which slows down all zones that access main memory more often, such as the monitoring zone.
Therefore, we estimate that the overhead might be even lower on production-ready devices and will only be influenced by the detection's complexity.

\section{Related Work} \label{r5detect:related-work}
The existing works most related to \PROJNAME{} are implementations of secure monitoring for mostly other architectures, such as ARM and x86.
Only a few papers focus on \ac{CFI} and RISC-V, and most of these require an extension of the \ac{ISA}~\cite{DBLP:conf/date/DeBGJ19, DBLP:conf/date/DelshadtehraniC21, DBLP:journals/sensors/ParkKKK22, DBLP:conf/dsd/ZgheibPRD22, DBLP:conf/eurosec/WangWYNZ22}.

Regarding \ac{TEE} on RISC-V, numerous designs and implementations are proposed, which are based again on non-standard extensions of the \ac{ISA}~\cite{DBLP:conf/uss/NoormanADSHHPVP13, DBLP:conf/uss/CostanLD16, DBLP:conf/ndss/WeiserWBMMS19}.
To our knowledge, only the open-source Keystone Enclave~\cite{DBLP:conf/eurosys/LeeKSAS20} and proprietary MultiZone\footnote{\url{https://hex-five.com/multizone-security-tee-riscv/}} adhere to the most recent RISC-V \ac{ISA} standards. %
There have also been recent efforts to bring \ac{TEE} features to standard RISC-V architectures~\cite{DBLP:conf/IEEEares/Bove22}.

Our Shadow Stack is also based on some previous works in the field.
In \cite{DBLP:journals/tissec/AbadiBEL09} the authors present their \ac{CFI} methods using a user-level shadow stack implementation and dynamic memory checks to achieve protection.
Subsequent works build on this idea to monitor bare-metal apps and smaller real-time operating systems~\cite{DBLP:conf/ecrts/WallsBBSOW19}.
With RIMI~\cite{DBLP:conf/iccad/KimLPAKK20}, the authors propose an instruction-level memory isolation RISC-V extension, which can then be used to protect a shadow stack.
Currently, there are efforts to implement and ratify a new extension (\enquote{Zisslpcfi}) for \ac{CFI} that includes a shadow stack~\cite{zisslpcfi2022}.

The topic of \aclp{HPC} has been ex\-ten\-sive\-ly re\-searched for different use cases, including security~\cite{DBLP:conf/sp/DasWAPM19}.
More specifically, \acp{HPC} are used in malware detection~\cite{DBLP:conf/isca/DemmeMSTWSS13, DBLP:journals/taco/WangCILK16, DBLP:conf/ccs/ZhouGJEJ18}, side-channel attack detection~\cite{DBLP:conf/isca/NomaniS15}, kernel rootkit detection~\cite{DBLP:conf/ccs/SinghEERC17} and more.
There are also approaches combining \acp{HPC} with \acp{TEE}, such as ARM TrustZone~\cite{DBLP:conf/sbesc/JungERSHM22} or \acl{SGX}~\cite{DBLP:conf/nordsec/LantzBA22}.
There are no works that implement universal \ac{HPC} monitoring on standard RISC-V architectures.

\section{Discussion} \label{r5detect:discussion}
In this section, we are discussing limitations and problems raised by related work (see \autoref{r5detect:related-work}) regarding our work on \PROJNAME{}.
In our implementation, we focused on two different detection methods, each with their own implications.

\subsection{CFI}
\ac{CFI} monitoring is limited by the target's program structure.
Small functions may cause problems, as there might not be enough space to introduce our label.
Therefore, not every function may be able to use the \ac{CFI} protection.
In addition, our current Shadow Stack implementation is static, which means that its size and the number of stacks must be known at compile time.
This is not much of a problem for our target low-power \ac{IoT} devices, as they are usually already limited in processing power and memory space, and developers need to account for the size and execution requirements of their applications anyway.
Therefore, a dynamic implementation would not significantly benefit our detection and would require more code and resources to implement.
For more powerful devices, that also may use a more complex Unix-based \ac{OS}, a dynamic approach might be preferable.

We did not consider the power consumption of devices, which was out-of-scope for our experiments.
There may be use cases where devices are attached to batteries instead of a power supply, but this sparks a number of other security considerations that our threat model does not cover.
Our goal is to show the potential of low-power hardware for security applications, regardless of energy consumption.

\subsection{HPC}
Our \ac{HPC} monitoring is mainly limited by the number of available counters on the hardware device.
In our case, we could choose from 23 different events, but we found that some events inherently depend on the execution environment and might be unsuited for some applications.
In future work, it would be interesting to find out what and how many real-world devices implement \acp{HPC} to find the sweet spot between hardware requirements and suitable event selection.

In previous work on the x86 and ARM architectures, two main issues with \acp{HPC} have been observed: \textit{overcounting} and \textit{non-deter\-mi\-nism}~\cite{DBLP:conf/ispass/WeaverTM13}.
The former describes the problem of incorrect performance counter values observed on some processors and architectures.
The latter issue concerns the tools, techniques, and methods used to obtain \ac{HPC} values, which may yield very different results even in a \enquote{strictly controlled environment for the same application}~\cite{DBLP:conf/sp/DasWAPM19}.
Especially these two issues evoke doubts on the effectiveness of performance counters for security applications.

We addressed these potential problems early in our design.
Non-determinism, coupled with race conditions between different processor cores and threads, is mostly excluded in our use case.
Our design is based on the MultiZone architecture, and our implementation features an own \textit{round-robin scheduler}, which cleanly separates different zone executions from each other.
Since we collect \ac{HPC} values per zone, it is impossible that other zones may interfere with the currently active zone and \acp{HPC}.
Therefore, we prevented the problem of non-determinism early in the design, which works regardless of the presence of multiple cores.

For the problem with overcounting, we unfortunately could not reproduce the problems described in previous works in our validation and evaluation experiments.
Being a pretty new field of research, there is a lack of research about the problem of overcounting for RISC-V.
We agree that security should not be solely based on \ac{HPC} measurements, but we disagree with works that dismiss them completely.
We see overcounting as a problem that might be based on individual processor implementations, but can be circumvented with better detection strategies, such as using a \enquote{moving average} approach to measurements instead of using fixed thresholds.
Nonetheless, further research is needed to assess the scale of this problem on RISC-V.

We observed an additional challenge to consider for \ac{HPC} monitoring on RISC-V: interrupts.
When interrupts are triggered (e.g., by sensors or other peripherals), the \acl{SM} calls the registered interrupt handler.
If these handlers are not considered during training and not included in the characteristic signature of a program, it may trigger a false positive in our detection, as the handler will influence the selected \ac{HPC} differently than the main program.
Since interrupts are triggered at unpredictable intervals all over the system, one needs to count and manage all interrupts in the \acl{SM} in order to even out the detection.
In our opinion, this is an often overlooked problem with \ac{HPC} measurements, and we did not find any mentions of this in related work.

\section{Conclusion} \label{r5detect:conclusion}
In this work we present \PROJNAME{}, a software-based \ac{CFI} detection with hardware-based security measures.
We implemented \PROJNAME{} on top of standard RISC-V features for \ac{IoT} and embedded devices that lack proper security features.
With our Shadow Stack approach, the majority of applications become resilient to \ac{CFI} and \ac{ROP} attacks at the cost of below 5~\% runtime overhead and with minimal ROM size requirements.
We expect this number to decrease significantly in the future, as Shadow Stacks might become a standard extension of the RISC-V ISA, shifting the solution to the hardware-accelerated micro-architectural implementation.

We combine the above protection with a detection based on \aclp{HPC}.
While severely limited on our test device, \acp{HPC} may become a significant building block for detecting unexpected and unwanted software behavior.
We also address common problems related to security based on \acp{HPC} in our discussion.
Combined with \ac{TEE}-related features, \PROJNAME{} can protect itself from a full compromise of the \ac{OS} and its user apps.

With our work, we hope to inspire more research in the field of \ac{IoT} and embedded security, as these devices have become ubiquitous and deserve more attention from the security community.

\bibliographystyle{splncs04}
\bibliography{references}

\end{document}